\documentclass[aps,twocolumn,groupedaddress,showpacs]{revtex4}
\usepackage{graphicx}
\begin{document}
\bibliographystyle{apsrev}


\title{Superconductivity in the Correlated Pyrochlore Cd$_2$Re$_2$O$_7$}


\author{R. Jin$^1$}
\email[]{email address: jinr@ornl.gov}
\author{J. He$^{2,1}$}
\author{S. McCall$^3$}
\author{C.S. Alexander$^3$}
\author{F. Drymiotis$^3$}
\author{D. Mandrus$^{1,2}$}
\affiliation{$^1$Solid State Division, Oak Ridge
National Laboratory, Oak Ridge, TN 37831}
\affiliation{$^2$Department of Physics and Astronomy, The
University of Tennessee, Knoxville, TN 37996}
\affiliation{$^3$National High Magnetic Field Laboratory and
Physics Department, Florida State University, Tallahassee,
FL32306}


\date{\today}

\begin{abstract}
We report the observation of superconductivity in high-quality
Cd$_2$Re$_2$O$_7$ single crystals with room-temperature pyrochlore
structure. Resistivity and ac susceptibility measurements
establish an onset transition temperature T$_c^{onset}$ = 1.47 K
with transition width $\Delta$T$_c$ = 0.25 K. In applied magnetic
field, the resistive transition shows a type-II character, with an
approximately linear temperature-dependence of the upper critical
field H$_{c2}$. The bulk nature of the superconductivity is
confirmed by the specific heat jump with $\Delta$C = 37.9
mJ/mol-K. Using the $\gamma$ value extracted from normal-state
specific heat data, we obtain $\Delta$C/$\gamma$T$_c$ = 1.29,
close to the weak coupling BCS value. In the normal state, a
negative Hall coefficient below 100 K suggests electron-like
conduction in this material. The resistivity exhibits a quadratic
T-dependence between 2 and 60 K, $\it{i.e.}$, $\rho
=\rho_0$+AT$^2$, indicative of Fermi-liquid behavior. The values
of the Kadowaki-Woods ratio A/$\gamma^2$ and the Wilson ratio are
comparable to that for strongly correlated materials.

\end{abstract}
\pacs{74.10.+v, 74.25.Fy, 74.25.Ha, 74.60.Ec, 71.27.+a}

\maketitle

Interest in oxide superconductors has been greatly stimulated by
the high critical temperatures (T$_c$) of the cuprates and the
unconventional superconductivity in Sr$_2$RuO$_4$. These materials
form in perovskite-like structures, where CuO$_2$ or RuO$_2$
layers play important roles in the occurrence of
superconductivity. Oxide superconductors with non-perovskite
structures are rare. In particular, while many oxides crystallize
in a pyrochlore structure with the general formula A$_2$B$_2$O$_7$
(where A and B are cations), no superconductivity has been
reported in the literature. At present, it is not clear why the
pyrochlore structure is unfavorable for superconductivity.
Previous studies indicate that the pyrochlores, like the spinels,
are geometrically frustrated \cite{subr}. The effect of geometric
frustration on the physical properties of spinel materials is
drastic, resulting in, for instance, heavy-fermion behavior in
LiV$_2$O$_4$ \cite{kondo}. To understand the role of geometrical
frustration in pyrochlores, we have investigated transport,
magnetic and thermodynamic properties of Cd$_2$Re$_2$O$_7$, the
only-known pyrochlore superconductor discovered two weeks ago
\cite{sakai}.

Although Cd$_2$Re$_2$O$_7$ was first synthesized in 1965
\cite{donohue}, its physical properties remained almost unstudied
except for specific heat measurements below 20 K \cite{black}.
Careful measurements of electrical resistivity, Hall effect,
specific heat and magnetic susceptibility of Cd$_2$Re$_2$O$_7$
single crystals indicate that there are at least two phase
transitions below room temperature: one near 200 K
\cite{jin1,jin2} and another around 1.5 K. In this communication,
we focus on the latter one. Both resistivity and ac susceptibility
indicate a superconducting transition at T$_c$ = 1.47 K. The
superconducting critical field, obtained from the resistive
transition, reveals an approximately linear temperature
dependence. Associated with the superconducting transition, the
specific heat exhibits a peak with jump $\Delta$C = 37.9 mJ/mol-K.
Above T$_c$, the Hall coefficient is negative, reflecting electron
dominated conduction. Both resistivity and Hall angle data exhibit
a T$^2$-dependence when approaching T$_c$ from high temperatures.
The T$^2$-behavior of the resistivity and the values of the
Kadowaki-Woods ratio A/$\gamma^2$ and the Wilson ratio suggest
that the ground state of Cd$_2$Re$_2$O$_7$ is a correlated Fermi
liquid.

Single crystals of Cd$_2$Re$_2$O$_7$ used in this study were grown
using a vapor-transport method with details described elsewhere
\cite{he}. The Cd:Re ratio was confirmed using electron microprobe
analysis, but no attempt was made to determine the oxygen content.
A previous study on crystals prepared by the same method claimed
an oxygen stoichiometry of 7 \cite{donohue}. The X-ray refinement
results confirm the pyrochlore structure with unit cell parameter
a = 10.2244(6) {\AA} at room temperature. This value is in
agreement with that obtained in Ref. \cite{donohue}. As pointed
out in Ref. \cite{jin2}, there is a subtle structure change at low
temperatures.

Fig.\ 1 shows the temperature dependence of the ac susceptibility
from a Cd$_2$Re$_2$O$_7$ single crystal, performed by using a
mutual inductance technique at an applied field of H $\sim$ 1 Oe
and a frequency of f = 1 kHz. The real part, $\chi$', reveals a
large diamagnetic signal below 1.15 K, marking the superconducting
transition. Below 0.75 K, $\chi$' is flat, indicating that the
superconducting transition is complete. We noticed that $\chi$'
was not saturated down to 0.3 K in polycrystalline
Cd$_2$Re$_2$O$_7$ \cite{sakai}.

\begin{figure}
\includegraphics[keepaspectratio=true, totalheight = 2.0 in, width = 2.0 in]{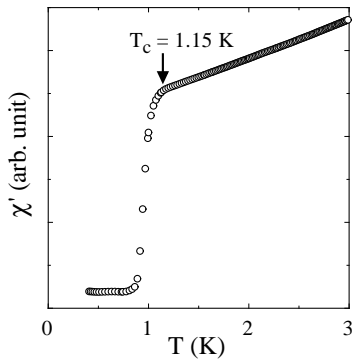}
\caption{Temperature dependence of the ac susceptibility (real
part) of a Cd$_2$Re$_2$O$_7$ single crystal.}
\end{figure}

Using a standard four-probe method, the dc electrical resistivity
of Cd$_2$Re$_2$O$_7$ has been investigated. Shown in Fig.\ 2 is
the temperature dependence of the electrical resistivity $\rho$
between 0.3 and 10 K at zero magnetic field. Associated with the
diamagnetic transition, $\rho$ also departs from high-temperature
behavior at 1.5 K and decreases abruptly to zero at 1.15 K,
corresponding to the onset of diamagnetism. The resistivity varies
from 10\% to 90\% of the normal-state value $\rho_N$ over a range
of approximately 0.25 K.

\begin{figure}
\includegraphics[keepaspectratio=true, totalheight = 2.0 in, width = 2.0 in]{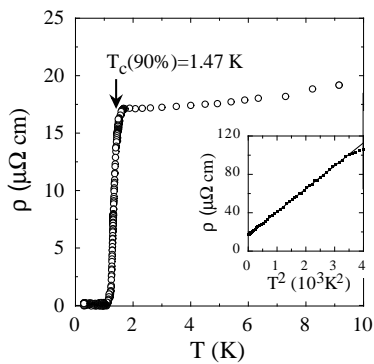}
\caption{Temperature dependence of the resistivity of
Cd$_2$Re$_2$O$_7$. The superconducting transition is indicated.
The inset shows the resistivity curve between 2 and 64 K plotted
as $\rho$ vs. T$^2$. The solid line is a fit to experimental data
between 2 and 60 K using $\rho =\rho_0$+AT$^2$.}
\end{figure}

Both resistivity and ac susceptibility measurements establish a
superconducting transition with the onset transition temperature
T$_c^{onset}$ = 1.47 K and a transition width $\Delta$T$_c$ =
T$_c$(90\%)-T$_c$(10\%) = 0.25 K for our Cd$_2$Re$_2$O$_7$ single
crystals, confirming the recent discovery \cite{sakai}. As
illustrated in the inset of Fig.\ 3, by applying a magnetic field
H perpendicular to the current I (H$\perp$I), the resistive
transition shifts to lower temperatures. The transition width
becomes wider with increasing H, a characteristic of type-II
superconductivity. We may define a resistive transition
temperature T$_c$(H) which satisfies the condition that
$\rho$(T$_c$,H) equals to a fixed percentage p of the normal-state
value $\rho_N$ for each field H. The values of T$_c$(H) for p =
10\%, 50\% and 90\% are shown in the main frame of Fig.\ 3,
represented by the upper critical field H$_{c2}$(T). In all cases,
we find that H$_{c2}$(T) depends more or less linearly on T with
no sign of saturation down to 0.3 K (see the solid lines). The
slope dH$_{c2}$/dT$\mid_{T=T_c}$ = -0.56T/K for p = 10\%, -0.62
T/K for p = 50\% and -0.83 T/K for p = 90\%. In the conventional
BCS picture, H$_{c2}$ is linear in T near T$_{c0}$ and saturates
in 0 K limit. Deviation may occur in the presence of strong
impurity scattering \cite {maeka}. In this case, the
Werthamer-Helfand-Hohenberg (WHH) formula \cite{whh} is often used
to describe the temperature dependence of H$_{c2}$ with
H$_{c2}$(0) = -0.693T$_c$(dH$_{c2}$/dT)$\mid_{T=T_c}$. The dashed
lines in Fig.\ 3 are the results of fitting H$_{c2}$(T) to the WHH
formula, yielding H$_{c2}^{WHH}$(0) = 0.48 T for p = 10\%, 0.57 T
for p = 50\% and 0.84 T for p = 90\%. Note that at lower
temperatures H$_{c2}$(T) no longer follows the WHH expression,
particularly for p = 10\% and 50\%. This suggests that the actual
H$_{c2}$(0) is larger than H$_{c2}^{WHH}$(0). To find out whether
H$_{c2}$(T) continuously increases in a linear fashion or
eventually saturates, resistivity measurements at lower
temperatures and higher magnetic fields are in progress.
Nevertheless, assuming H$_{c2}$(0) = H$_{c2}^{WHH}$(0), we may
estimate the superconducting coherence length $\xi_{GL}$ using
Ginzburg-Landau formula $\xi_{GL}$ =
($\Phi_0$/2$\pi$H$_{c2}$)$^{1/2}$,
 where $\Phi_0$ = 2.07$\times$10$^{-7}$ Oe cm$^2$. This results in the zero-temperature coherence length
$\xi_{GL}$(0) = 263 {\AA} for p = 10\%, 240 {\AA} for p = 50\% and
198 {\AA} for p = 90\%.

\begin{figure}
\includegraphics[keepaspectratio=true, totalheight = 2.0 in, width = 3.0 in]{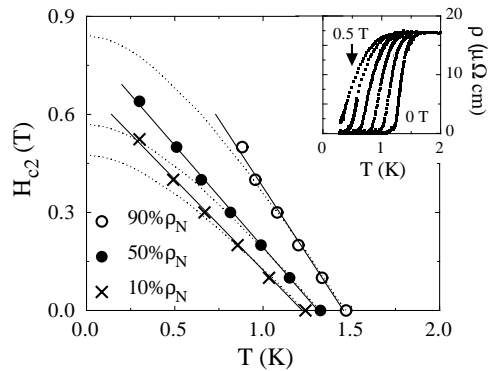}
\caption{Temperature dependence of the upper critical field
H$_{c2}$ deduced from the resistivity measured at 90\% (open
circles), 50\% (solid circles) and 10\% (cross) of the
normal-state value $\rho_N$. The dash lines represent the WHH
approach (see the text). The solid lines are the linear fit to
experimental H$_{c2}$(T). The inset shows the temperature
dependence of resistivity at H = 0, 0.1, 0.2, 0.3, 0.4, and 0.5
T.}
\end{figure}

Is Cd$_2$Re$_2$O$_7$ indeed a dirty superconductor? To address
this issue, we need to compare the mean free path {\it l} with
Pippard coherence length $\xi_0 = \hbar$v$_F$/$\pi\Delta$(0),
where v$_F$ is the Fermi velocity and the zero-temperature energy
gap $\Delta$(0) = 1.764k$_B$T$_c$ according to the BCS theory.
Information about {\it l} and $\xi_0$ may be obtained from Drude
relation {\it l} =
$\hbar$(3$\pi^2$)$^{1/3}$/e$^2$$\rho_0$n$^{2/3}$ and
$\xi_0=\hbar^2$(3$\pi^2$n)$^{1/3}$/1.764$\pi$mk$_B$T$_c$, where m
is the electron rest mass, n is the carrier density and $\rho_0$
is the residual resistivity at 0 K at which the scattering is
essentially from impurities. In analyzing the data above T$_c$, we
found that the resistivity exhibits a quadratic temperature
dependence over a wide temperature regime. Shown in the inset of
Fig.\ 2 is the plot of $\rho$ vs. T$^2$ between 2 and 64 K. Note
that $\rho$ varies approximately linearly with T$^2$ below ~ 60 K.
By fitting the resistivity data between 2 and 60 K using a formula
$\rho = \rho_0$+AT$^2$, we obtain the residual resistivity
$\rho_0$ = 17 $\mu\Omega$ cm and constant A = 0.024 $\mu\Omega$
cm/K$^2$. As illustrated in the inset of Fig.\ 2 by the solid
line, the above formula fits the experimental data very well. This
indicates the importance of the Umklapp process of the
electron-electron scattering at low temperatures and is consistent
with the formation of a Fermi liquid state. The extrapolated
residual resistivity of our crystals is approximately 235 times
lower than that for polycrystals \cite{sakai}, reflecting a much
lower level of impurities in our single crystals. Interestingly,
there is little difference in T$_c$. We recall that the
superconductivity in Sr$_2$RuO$_4$ is completely suppressed as
$\rho_0$ exceeds $\sim$ 1 $\mu\Omega$ cm \cite{mackenzie}. This
suggests that the impurity effect on superconductivity in
Cd$_2$Re$_2$O$_7$ is much weaker than in Sr$_2$RuO$_4$. To assess
the carrier density n, we have performed Hall effect measurements.
Using the standard four-point technique, the Hall component was
derived from the antisymmetric part of the transverse resistivity
under magnetic field reversal at a given temperature. As displayed
in Fig.\ 4, the Hall coefficient R$_H$ is T-dependent and has
negative sign below 100 K. This suggests that the Fermi surface of
Cd$_2$Re$_2$O$_7$ may contain several sheets and electrons
dominate the electrical conduction. Nevertheless, the T$^2$
behavior of the Hall angle cot$\theta_H$ (see the inset of Fig.\
4) at low temperatures suggest that both longitudinal and
transverse transport properties are controlled by the same
scattering, unlike the high-T$_c$ cuprate materials
\cite{anderson}. Using the simple Drude relation, we estimate n =
-1/eR$_H$ $\sim$ 7$\times$10$^{21}$ cm$^{-3}$ for T = 5 K. (We are
aware that the simple Drude relation may not hold if the Fermi
surface of Cd$_2$Re$_2$O$_7$ consists of multibands.) Inserting
the estimated n and $\rho_0$, we obtain {\it l} $\sim$ 204 {\AA}
and $\xi_0$ $\sim$ 6365 {\AA}. Since $\xi_0$ is much larger than
{\it l}, Cd$_2$Re$_2$O$_7$ is in the dirty limit.

\begin{figure}
\includegraphics[keepaspectratio=true, totalheight = 2.0 in, width = 3.0 in]{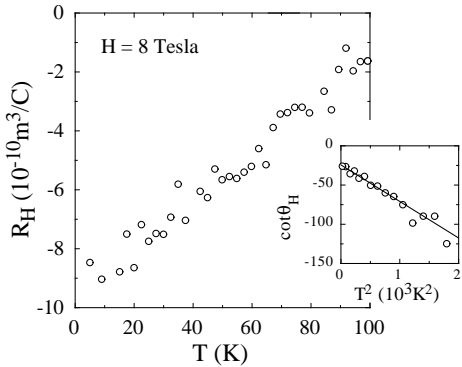}
\caption{Temperature dependence of Hall coefficient R$_H$ between
5 and 100 K. The inset shows Hall angle cot$\theta$$_H$ vs. T$^2$
(open circles). The solid line is the linear fit of experimental
data between 2 and 40 K.}
\end{figure}

Given the values of {\it l} and $\xi_0$, we may also estimate the
GL coherence length $\xi_{GL}$(0) using $\xi_{GL}$(0) $\sim$
0.855($\xi_0${\it l})$^{1/2}$ for dirty superconductors
\cite{tinkham}. This relation yields $\xi_{GL}$(0) $\sim$ 927
{\AA}, a few times larger than that obtained from
H$_{c2}^{WHH}$(0). There could be several reasons to cause the
discrepancy. One possibility is that the slope
dH$_{c2}$/dT$\mid_{T=T_c}$ is unexpectedly large, which results in
large H$_{c2}^{WHH}$(0) and consequently small $\xi_{GL}$(0). As
mentioned in Ref. \cite{sakai}, the large value of
dH$_{c2}$/dT$\mid_{T=T_c}$ may imply that the Cooper pairs are
composed of heavy quasiparticles since it is proportional to the
effective mass m* \cite{tinkham}. Given the fact that the
effective electron mass is significantly enhanced due to geometric
frustration in LiV$_2$O$_4$ \cite{kondo}, it is not surprising
that such effect also plays a similar role in Cd$_2$Re$_2$O$_7$.
Further evidence for heavy quasiparticles can be found from
specific heat data.

The specific heat of Cd$_2$Re$_2$O$_7$ was measured using a
relaxation calorimeter, where the contribution from the addenda
has been carefully subtracted. Fig.\ 5 shows the temperature
dependence of specific heat between 0.4 and 2.0 K. Note that the
specific heat reveals a pronounced peak associated with the
superconducting transition, confirming the bulk nature of the
superconductivity. At the midpoint of the transition T$_c^{mid}$ =
0.99 K, we determine the specific heat jump $\Delta$C = 37.9
mJ/mol-K. In the weak coupling limit, $\Delta$C is expected to
approach 1.43$\gamma$T$_c$ \cite{tinkham}, where $\gamma$ is the
Sommerfeld coefficient and can be obtained from the normal-state
specific heat. An expression of the form C =
$\gamma$T+$\beta$T$^3$ is usually used to describe the specific
heat data at temperatures well below the Debye temperature
$\Theta_D$, {\it i}.{\it e}., T $\ll \Theta_D$, where $\beta$ =
N(12/5)$\pi^4$R$\Theta_D^{-3}$, R = 8.314 J/mol-K and N = 11 for
Cd$_2$Re$_2$O$_7$. The T-term comes from the electronic
contribution (C$_e$) and the T$^3$-term arises from the lattice
contribution (C$_l$). By plotting our specific heat data as C/T
vs. T$^2$ as shown in the inset of Fig.\ 5, a linearity is clearly
seen below $\sim$ 6.2 K. We fit the data between 1.2 and 6.2 K
using the above formula and obtain $\gamma$ = 29.6 mJ/mol-K$^2$
and $\Theta_D$ = 397 K, slightly higher than those given in Ref.
\cite{black}. This leads that $\Delta$C/$\gamma$T$_c^{mid}$ =
1.29, close to the weak coupling BCS result. In the framework of
the BCS theory, the superconducting-state electronic specific heat
C$_{es}$(T) is expected to decay exponentially with T, {\it
i}.{\it e}., C$_{es}$ = ae$^{-b/T}$ (a and b are T-independent
constants). As can be seen in Fig.\ 5, the BCS formula (dashed
line) describes our experimental data very well down to 0.4 K with
a = 0.307 J/mol-K and b = 1.52 K. However, in the absence of the
data at lower temperatures (T $\le$ 0.3 K), it is not clear
whether or not Cd$_2$Re$_2$O$_7$ is a BCS-type superconductor.

In comparison with other pyrochlores \cite{subr,ishii,mandrus},
the $\gamma$ value for Cd$_2$Re$_2$O$_7$ is large. A large
electronic specific heat at low temperatures is usually observed
in strongly correlated Fermi-liquid systems like heavy-fermion
materials \cite{kondo,kw,miyaka,takimoto} and
Sr$_{n+1}$Ru$_n$O$_{3n+1}$ series \cite{ikeda,ikeda2} due to an
effective mass enhancement. It is known that for such systems, the
Kadowaki-Woods ratio A/$\gamma^2$ is expected to approach the
universal value A/$\gamma^2$ = 1.0$\times$10$^{-5}$ $\mu\Omega$
cm/(mJ/mol-K)$^2$ \cite{kw,miyaka,takimoto}. For
Cd$_2$Re$_2$O$_7$, we obtain A/$\gamma^2$ = 2.7$\times$10$^{-5}$
$\mu\Omega$ cm/(mJ/mol-K)$^2$, very close to that found for the
heavy fermion compound UBe$_{13}$. The consistency of A/$\gamma^2$
value with the universal description suggests that the electrons
are strongly correlated in Cd$_2$Re$_2$O$_7$. In such a system,
the Wilson ratio R$_W = \pi^2k_B^2\chi_{spin}/3\mu_B^2\gamma$ is
expected to be greater than one, where $\chi_{spin}$ denotes the
spin susceptibility, k$_B$ is the Boltzmann's constant and $\mu_B$
is the Bohr magneton. According to our dc susceptibility data
presented in Ref. \cite{jin2}, we estimate that $\chi_{spin}$ =
4.6$\times$10$^{-4}$ emu/mol at low temperatures. This gives that
R$_W \sim$ 1.3, well exceeding the unity value for a free electron
system.

\begin{figure}
\includegraphics[keepaspectratio=true, totalheight = 2.0 in, width = 3.0 in]{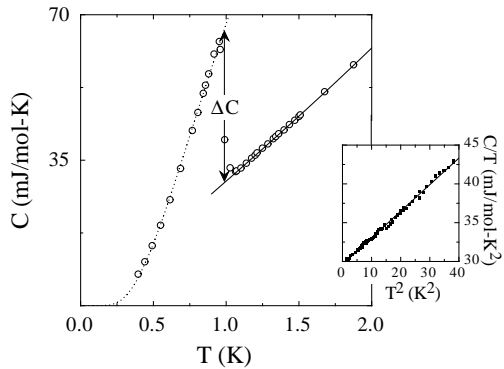}
\caption{Temperature dependence of the specific heat of
Cd$_2$Re$_2$O$_7$ between 0.3 and 2.0 K. The dash line is the fit
of experimental data to C = ae$^{-b/T}$+$\beta$T$^3$ using $\beta$
value extracted from the normal-state specific heat data. The
inset shows specific heat C versus temperature above T$_c$ plotted
as C/T against T$^2$. The solid lines (in both the inset and main
panel) are the fit of experimental data to C =
$\gamma$T+$\beta$T$^3$ between 1.2 and 6.2 K.}
\end{figure}

In summary, from transport and ac magnetic susceptibility
measurements, we confirm the superconducting transition with
T$_c^{onset}$ = 1.47 K in Cd$_2$Re$_2$O$_7$ single crystals. The
bulk nature of the superconductivity has been confirmed by the
specific heat jump across T$_c$. The ratio
$\Delta$C/$\gamma$T$_c^{mid}$ is close to the weak coupling BCS
value. However, the almost linear temperature dependence of
resistive critical field cannot be described by the WHH formula
for a dirty superconductor. The T$^2$ dependence of the
resistivity, the large values of dH$_{c2}$/dT$\mid_{T=T_c}$ and
the Wilson ratio, and the value of A/$\gamma^2$, all suggest that
the electrons in Cd$_2$Re$_2$O$_7$ are strongly correlated with
the enhanced effective mass, resulting possibly from geometric
frustration.


\begin{acknowledgments}
We would like to thank Dr. J.E. Crow for arranging the low
temperature experiments and fruitful discussions. We are also
grateful to Dr. B.C. Sales for useful discussions and comments.
Oak Ridge National laboratory is managed by UT-Battelle, LLC, for
the U.S. Department of Energy under contract DE-AC05-00OR22725.
\end{acknowledgments}

\bibliography{Cd2Re2O7bib.tex}

%
%

%
%

\end{document}